\documentclass[12pt]{article}

\usepackage{sbc-template}
\usepackage{graphicx,url}
\usepackage[utf8]{inputenc}
\usepackage[english]{babel}
\usepackage{booktabs}
\usepackage{array}
\usepackage{xcolor}

\title{Legal Nugget Extraction for Granular Retrieval over Long Jurisprudential Texts}

\author{Lucas Pereira\inst{1}, Erick Brito\inst{2}, Roberto Lotufo\inst{3},
Jayr Pereira\inst{2}}

\address{Tribunal de Contas da Uni\~ao (TCU)
\nextinstitute
Universidade Federal do Cariri (UFCA)
\nextinstitute
NeuralMind.ai
\\
\email{jayr.pereira@ufca.edu.br}
}

\begin{document}

\maketitle

\begin{abstract}
Legal retrieval over jurisprudential collections is challenging because court
decisions are long, heterogeneous documents whose relevant legal thesis may
occupy only a small portion of the text. This paper asks whether legal nuggets,
defined as short and self-contained legal theses extracted from source
documents, can improve dense retrieval over Brazilian legal collections. We
propose a pipeline that extracts nuggets from each document, indexes them with
embeddings, retrieves nugget-level evidence, and aggregates the retrieved
nuggets back to document-level rankings. We evaluate this approach on four
Portuguese legal retrieval benchmarks from the JUA ecosystem, reporting
NDCG@10, MAP@10, and MRR@10. Nugget retrieval substantially improves the two
jurisprudential datasets: on JUA-Juris, NDCG@10 increases from 0.10265 to
0.20461, and on JurisTCU from 0.20898 to 0.32696. However, it underperforms
full-document retrieval on NormasTCU and BR-TaxQA, and an embedding-model
ablation shows that strong domain-adapted retrievers can remain better in the
full-document setting. The results demonstrate that legal nuggets can be useful
for jurisprudence search, especially when queries are formulated as legal
theses, but they may not transfer equally well to other legal retrieval
scenarios.
\end{abstract}

\section{Introduction}

Searching jurisprudence is a central task in legal practice because lawyers,
judges, auditors, and researchers often need to find prior decisions that
support, distinguish, or contradict a specific legal thesis. Unlike many web
documents, jurisprudential texts are long, formal, and argumentative: a single
ruling may contain procedural history, factual background, references to
parties, multiple legal issues, and one or more generalizable holdings. As a
result, the document that should be retrieved is often relevant because of a
small thesis-bearing fragment rather than because of its full textual content.
Dense retrieval has become a central paradigm for modern information retrieval
and question answering \cite{karpukhin2020dpr,thakur2021beir}, but embedding a
complete jurisprudential document into a single vector may dilute the signal of
the legal thesis that matters for the query.

The idea of retrieving smaller information units has appeared in several
contexts. Passage retrieval and retrieval-augmented generation commonly
decompose long documents into retrievable chunks \cite{lewis2020rag}, while
late-interaction models such as ColBERT preserve more granular matching
signals inside the retrieval model \cite{khattab2020colbert}. In question
answering and evidence-centered retrieval, short atomic pieces of information
can also serve as evidence units that are easier to rank, inspect, and combine
than full documents. Recent RAG work has made this explicit through
\emph{information nuggets}: minimal facts, claims, or question-answer units
used to evaluate answer completeness, preserve provenance, guide generation, and
construct retrieval benchmarks \cite{lajewska2025ginger,dietz2026nuggets,
thakur2025freshstack}. These approaches share a common intuition: retrieval
may improve when the indexed or evaluated unit is closer to the unit of
information requested by the query.

This paper explores legal nuggets as one possible granular representation for
jurisprudential search. We define a legal nugget as a short, self-contained
statement that captures an atomic legal argument or thesis consolidated by a
document. Instead of indexing the full document text directly, our method
extracts nuggets from each document, embeds the nuggets, searches over the
nugget index, and then maps the retrieved nuggets back to their original
documents. The motivation is that many jurisprudential queries resemble legal
theses rather than complete case narratives. If the indexed representation is
also thesis-like, retrieval may become more robust to noise. At the same time,
granular indexing introduces risks: extraction errors may remove important
context, and ranking several nuggets per document requires aggregation back to
the document level. This makes nugget retrieval a promising but empirical
hypothesis, not an automatic replacement for full-document retrieval.

In this context, our research question is: \emph{Can legal nuggets improve dense retrieval over
long jurisprudential texts in Brazilian legal collections?} For answering this
question, we compare a full-document dense retrieval baseline with a
nugget-based index in which each retrieved nugget is aggregated back to its
source document.

We evaluate this idea on four datasets from the JUA legal retrieval ecosystem
\cite{pereira2026juabenchmarkinformation}: JUA-Juris, JurisTCU
\cite{juristcu}, NormasTCU, and BR-TaxQA \cite{br-tax-qa}. The experiments
compare the nugget-based index against a full-document dense retrieval
baseline using the same embedding model. The results indicate a clear split.
Nugget retrieval substantially improves results on jurisprudential datasets,
especially when queries are ementas or legal theses. In contrast, it
underperforms full-document retrieval on NormasTCU and BR-TaxQA in terms of
NDCG@10 and MAP@10, although MRR@10 improves in NormasTCU. This mixed
behavior is consistent with prior observations that
legal NLP benefits from domain-aware representations, but that such
representations must be matched to the task and document genre
\cite{chalkidis2020legalbert}.

The contribution of this work is threefold: (i) we propose a simple pipeline
for legal nugget extraction and nugget-level dense retrieval; (ii) we provide
an empirical comparison across four Portuguese legal retrieval tasks; and
(iii) we analyze when granular legal representations help, arguing that the
benefit depends on the alignment between prompt, document genre, and query
type.

\section{Related Work}

Dense retrieval represents queries and documents as vectors and ranks
documents by vector similarity. Dual-encoder models such as DPR
\cite{karpukhin2020dpr} showed that dense retrieval can outperform sparse
term matching in open-domain question answering, while BEIR
\cite{thakur2021beir} highlighted the importance of evaluating retrieval
methods across heterogeneous tasks. Neural retrieval architectures have also
explored more granular forms of matching. ColBERT \cite{khattab2020colbert},
for example, performs late interaction over token-level embeddings rather than
compressing a whole document into a single vector.

Granularity is also central to retrieval-augmented generation, where systems
typically retrieve passages or chunks instead of entire documents
\cite{lewis2020rag}. Chunking reduces input length and can improve local
semantic matching, but generic fixed-size chunks may not correspond to
meaningful legal units. Legal texts are highly structured and argument-driven:
the most useful unit may be a legal thesis, rule application, or precedent
statement rather than an arbitrary window of tokens \cite{ma2024structural}.

Information nuggets offer a complementary granular representation for
retrieval and evaluation. Recent
work uses nuggets as minimal facts, claims, or question-answer units for
grounded generation and RAG evaluation. GINGER extracts, clusters, ranks, and
summarizes information nuggets to improve grounded response generation
\cite{lajewska2025ginger}, while Q\&A nuggets have been used to preserve
citation provenance and guide selection in RAG systems \cite{dietz2026nuggets}.
Nuggets have also been used to construct realistic retrieval benchmarks with
nugget-level support judgments \cite{thakur2025freshstack} and to compare
LLM-based relevance judgment methods \cite{arabzadeh2025llmjudgments}. In the
legal domain, Farzi et al. use factual nuggets to help legal professionals
compare and improve AI-generated summaries of legal depositions
\cite{farzi2026legaldepositions}. These works motivate nuggets as
interpretable evidence units, but they do not directly evaluate nugget indexing
for jurisprudential document retrieval.

Prior work on legal NLP has emphasized that legal language and legal tasks
require domain-aware representations \cite{chalkidis2020legalbert}. In legal
retrieval, this suggests that the document representation should reflect the
type of legal information being sought. Our work follows this intuition by
using large language models to extract legal theses from long jurisprudential
documents before embedding them for retrieval.

\section{Legal Nugget Retrieval}

\begin{figure}[t]
\centering
\includegraphics[width=\textwidth]{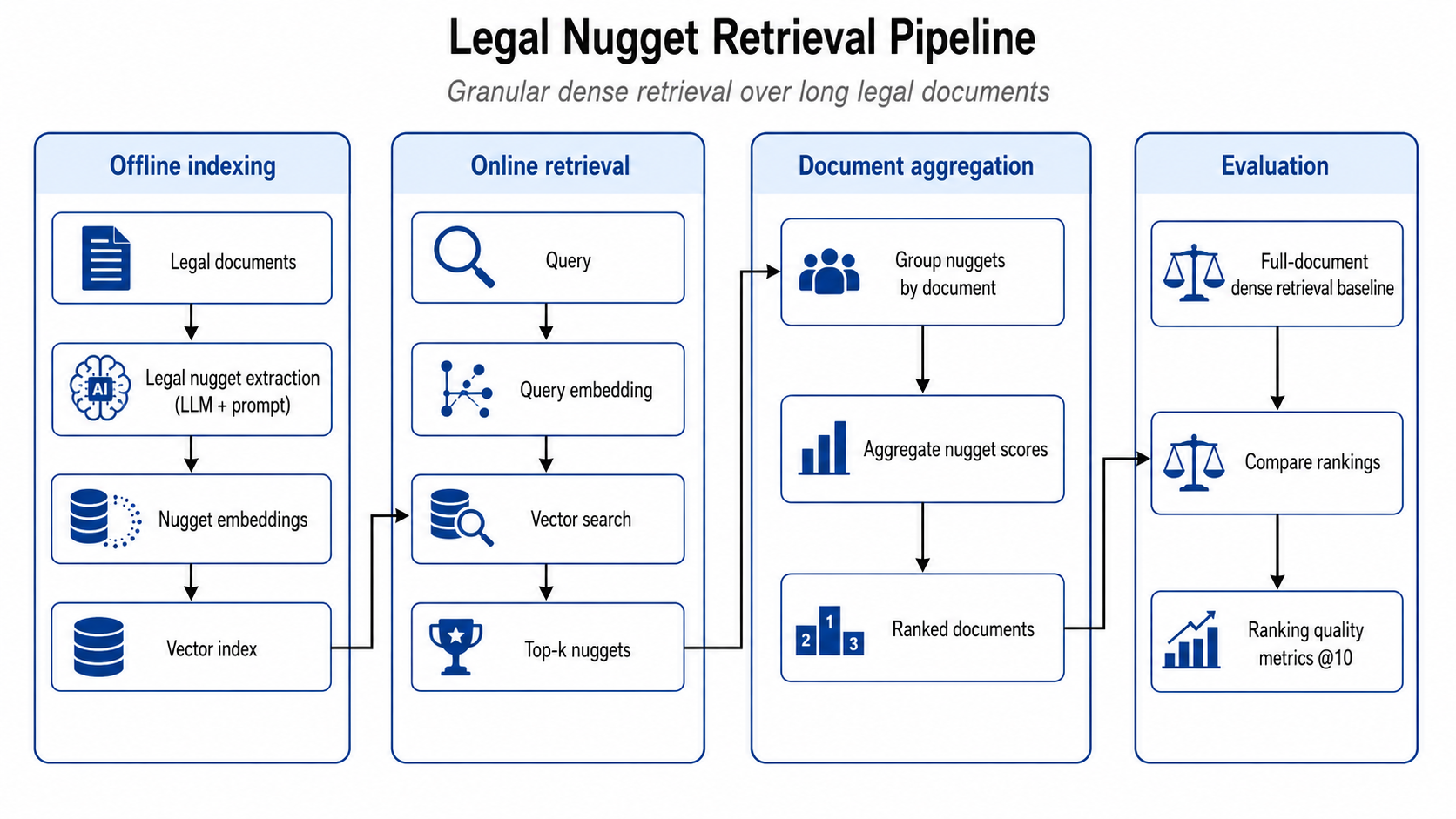}
\caption{Legal nugget retrieval pipeline. The offline phase converts legal
documents into an indexed nugget collection, while the online phase retrieves
nuggets for a query and aggregates them into a document ranking.}
\label{fig:pipeline}
\end{figure}

Figure~\ref{fig:pipeline} summarizes the proposed pipeline. The method has two
phases. In the offline phase, a corpus of jurisprudential documents is
converted into a nugget index: documents are passed to a legal nugget
generator, the generated nuggets are linked to their source document
identifiers, and an embedding model maps each nugget to a vector stored in a
vector index. In the online phase, a user query is embedded with the same
embedding model, the vector index retrieves the most similar nuggets, and a
document aggregator converts nugget-level evidence into a document-level
ranking.

\subsection{Nugget Extraction}

Nugget extraction is the step that turns a long legal document into a set of
smaller semantic units. In this work, a legal nugget is not merely a passage or
a fixed-size chunk. It is an atomic statement of legal content that can be
searched independently from the document that contains it. This follows recent
nugget-based work in retrieval and RAG, where compact information units are
used to represent factual evidence, guide selection, and preserve provenance
\cite{lajewska2025ginger,dietz2026nuggets}. For jurisprudence, we define
nuggets as reusable legal theses or arguments: compact statements that connect
a legal rule or principle, a relevant factual pattern, and the conclusion
consolidated by the decision.

This definition is motivated by the structure of jurisprudential documents.
Judicial and administrative rulings often contain procedural history,
identification of parties, factual background, citations, several legal issues,
and one or more holdings. A query, however, may target only one of these legal
holdings. Indexing the full document as a single unit can therefore dilute the
matching signal, especially when the relevant thesis is surrounded by long
context that is useful for legal interpretation but less useful for first-stage
retrieval.

The extraction step aims to preserve the part of the document that is most
likely to be reused in retrieval: the legal proposition itself. The prompt asks
the generator to avoid procedural details, isolated citations, party names,
document identifiers, purely factual descriptions, and generic dispositive
statements. The resulting nugget should be self-contained enough to be matched
against a query without requiring the full document, while still maintaining
its provenance through the identifier of the source document.
Figure~\ref{fig:prompt} shows a simplified version of the
jurisprudence-oriented prompt used to operationalize this definition.

\begin{figure}[ht]
\centering
\begingroup
\setlength{\fboxsep}{8pt}
\setlength{\fboxrule}{0.6pt}
\fcolorbox{gray!55}{gray!10}{%
\begin{minipage}{0.88\textwidth}
\small
You receive the text of a jurisprudential summary or ruling. Extract all
atomic legal arguments consolidated by the decision. Each nugget must express
one reusable legal thesis connecting a legal rule or principle, a factual
pattern, and a conclusion.\\[0.4em]
Return only legal arguments. Ignore procedural history, party names, document
identifiers, isolated citations, purely factual descriptions, and generic
dispositive statements. Each nugget must be self-contained and understandable
without the full document.\\[0.4em]
Return a valid JSON object containing a list of nugget texts. Each text is
later linked to the source document identifier by the indexing pipeline.
\end{minipage}%
}
\endgroup
\caption{Simplified version of the jurisprudential legal nugget extraction
prompt used for JUA-Juris and JurisTCU. Other datasets may use task-specific
prompts.}
\label{fig:prompt}
\end{figure}

\subsection{Indexing and Search}

Once extracted, nuggets become the retrievable units of the collection, in a
process similar to nugget-based retrieval and generation pipelines that use
atomic information units to guide selection and preserve provenance
\cite{dietz2026nuggets,lajewska2025ginger}. This changes the retrieval
problem from matching a query against complete legal documents to matching it
against candidate legal propositions. The intuition is that a query about a
legal thesis may be closer to one specific nugget than to the full decision
that contains it. In this setting, the document remains the unit of evaluation
and citation, but the nugget acts as the first-stage evidence used to find that
document.

Indexing nuggets therefore separates two roles that are conflated in standard
document retrieval. The nugget represents the semantic content to be matched,
while the source document represents the legal authority to be returned. This
distinction is important for long jurisprudential texts: a decision may contain
several legal propositions, and only one of them may correspond to the
information need expressed by the query. A nugget index allows these
propositions to compete individually before the system decides which documents
to return.

The final step maps nugget-level evidence back to document-level rankings. For
each query, the system first retrieves the top \(N\) nuggets from the vector
index, where \(N=10{,}000\) in our experiments. Each retrieved nugget has a
similarity score and a source document identifier. Documents are then
deduplicated by aggregating all retrieved nuggets that point to the same source
document. In the main experiments, we use a first-evidence aggregation
strategy: because retrieved nuggets are processed in decreasing similarity
order, the score assigned to a document is the score of its highest-ranked
retrieved nugget. Equivalently, for a query \(q\), document \(d\), retrieved
nuggets \(n_i\), scores \(s_i\), and source mapping \(doc(n_i)\), the document
score is
\(S(q,d)=\max\{s_i \mid doc(n_i)=d\}\) over the retrieved nugget set. The final
ranking is obtained by sorting documents by \(S(q,d)\), so each source
document appears at most once.

This aggregation choice treats a single highly relevant legal proposition as
sufficient evidence for retrieving its source document. We also implemented
alternative strategies, such as averaging the top \(k\) nugget scores from the
same document, which favor documents supported by multiple strong pieces of
evidence. These choices are part of the retrieval model, not merely
implementation details, because they determine how granular evidence is
translated into the document ranking expected by legal search benchmarks.

\section{Experimental Setup}

\subsection{Datasets}

We evaluate on four Portuguese legal retrieval datasets from the JUA benchmark
ecosystem \cite{pereira2026juabenchmarkinformation}:

\begin{itemize}
  \item \textbf{JUA-Juris}: a jurisprudential retrieval task introduced as
  part of JUA. The corpus is derived from curated TCU jurisprudence excerpts,
  and each query is based on an \textit{enunciado}, i.e., a concise statement
  of the legal thesis associated with a ruling. Relevance is binary and
  centered on the exact query--document pair, making this dataset especially
  close to thesis-to-jurisprudence retrieval.
  \item \textbf{JurisTCU}: a jurisprudential corpus from the Brazilian Federal
  Court of Accounts with query relevance judgments \cite{juristcu}. The
  original collection contains curated jurisprudence records with metadata and
  textual fields such as \texttt{ENUNCIADO} and \texttt{EXCERTO}, which
  correspond to the summary of a ruling and the supporting decision excerpt.
  Its queries include real keyword searches and synthetic queries derived from
  ruling summaries, with relevance judgments that were manually verified.
  \item \textbf{NormasTCU}: a retrieval task over normative documents from the
  same legal-administrative domain. Unlike jurisprudential collections, these
  documents are normative acts that establish general rules and are typically
  organized into provisions such as articles and paragraphs. In JUA, documents
  are represented by combining the subject summary field with the full text of
  the norm, and relevance is graded.
  \item \textbf{BR-TaxQA}: a tax-law question answering retrieval benchmark
  with references for Brazilian personal income tax law \cite{br-tax-qa}. It
  is based on FAQ-style questions and answers from the Brazilian Federal
  Revenue Service. The retrieval task is question-driven: systems must retrieve
  the answer or reference material associated with a taxpayer question, with
  graded relevance reflecting primary and secondary linked answers.
\end{itemize}

The main target of this study is jurisprudential retrieval. Accordingly,
JUA-Juris and JurisTCU use the jurisprudence-oriented nugget prompt described
in Section~3, which focuses on legal theses consolidated by rulings. NormasTCU
and BR-TaxQA are included as out-of-genre checks: they allow us to test
whether the same granular retrieval idea transfers beyond jurisprudence to
normative and question-driven legal search. For these two datasets, we use
slightly different prompts, adapted to extract atomic legal facts from
normative texts and tax question--answer material rather than jurisprudential
holdings.

\subsection{Models and Metrics}

Nuggets are generated with \texttt{gpt-5.4-nano}. This model is used only for
the extraction step: it reads each source document and produces the nugget
texts that are later indexed. Retrieval is performed by embedding the generated
nuggets and the queries with dense embedding models; the generator itself is
not used at search time. The implementation, run configurations, and prompts
are available in a public repository.\footnote{\url{https://github.com/ufca-llms/nuggets-search}}

The main comparison uses \texttt{text-embedding-3-small} for both the
full-document baseline and the nugget index. Baseline scores are taken from
the public JUA
leaderboard\footnote{\url{https://huggingface.co/spaces/ufca-llms/jua-leaderboard}},
which reports standardized results for the benchmark described by
Pereira et al. \cite{pereira2026juabenchmarkinformation}. The nugget runs use
the same embedding model, but the indexed units are extracted nuggets rather
than full documents. We use this proprietary embedding model as the main
experimental setting because it provides a public, standardized JU\'A
leaderboard baseline and allows us to isolate the effect of changing the
retrieval unit from complete documents to legal nuggets under the same encoder.
The open-weight models are therefore treated as an ablation: they test whether
the observed behavior persists when the embedding model changes, rather than
serving as the primary point of comparison. We index vectors with FAISS exact
search over L2-normalized embeddings. In the ablation, we additionally evaluate Qwen3
Embedding 4B and 8B \cite{zhang2025qwen3embedding}, as well as JUA-adapted
4B retrievers \cite{pereira2026domainadaptivedenseretrievalbrazilian}. These
choices instantiate the generic pipeline described in
Figure~\ref{fig:pipeline}; the pipeline itself does not depend on these
specific models.

We report standard retrieval metrics at cutoff 10: NDCG@10, a position-aware
metric that discounts lower-ranked hits and supports graded relevance
\cite{jarvelin2002cumulated}; MAP@10, which summarizes precision at ranks
where relevant documents appear \cite{manning2008introduction}; and MRR@10,
which emphasizes the rank of the first relevant hit
\cite{voorhees2000building}.

\section{Results and Discussion}

Table~\ref{tab:results} reports the main comparison across the four datasets.

\begin{table}[ht]
\centering
\caption{Main retrieval results comparing full-document indexing with legal
nugget indexing across four Brazilian legal datasets. Both approaches use the
same embedding model in each dataset.}
\label{tab:results}
\small
\setlength{\tabcolsep}{3pt}
\begin{tabular}{@{}llrrr@{}}
\toprule
Dataset & Approach & NDCG@10 & MAP@10 & MRR@10 \\
\midrule
JUA-Juris & Full document & 0.10265 & 0.07713 & 0.07732 \\
JUA-Juris & Legal nuggets & \textbf{0.20461} & \textbf{0.15756} & \textbf{0.15756} \\
\midrule
JurisTCU & Full document & 0.20898 & 0.08728 & 0.44645 \\
JurisTCU & Legal nuggets & \textbf{0.32696} & \textbf{0.17368} & \textbf{0.58299} \\
\midrule
NormasTCU & Full document & \textbf{0.26250} & \textbf{0.15815} & 0.42702 \\
NormasTCU & Legal nuggets & 0.24928 & 0.11562 & \textbf{0.49169} \\
\midrule
BR-TaxQA & Full document & \textbf{0.74126} & \textbf{0.65848} & \textbf{0.76473} \\
BR-TaxQA & Legal nuggets & 0.65566 & 0.56618 & 0.66025 \\
\bottomrule
\end{tabular}
\end{table}

The jurisprudential datasets show the strongest gains. On JUA-Juris,
nugget retrieval nearly doubles NDCG@10, from 0.10265 to 0.20461, and also
approximately doubles MAP@10 and MRR@10. On JurisTCU, NDCG@10 increases from
0.20898 to 0.32696, while MAP@10 almost doubles from 0.08728 to 0.17368.

The results are different for NormasTCU and BR-TaxQA. In NormasTCU, nuggets
slightly underperform full-document retrieval in NDCG@10 and MAP@10, but
improve MRR@10. This suggests that the nugget index may move the first
relevant result earlier while weakening the overall top-ranked ordering.
In BR-TaxQA, the full-document baseline remains stronger across all reported
metrics.

\subsection{Interpretation}

The three metrics highlight different retrieval behaviors. NDCG@10 is
position-aware and sensitive to graded relevance across the top ranks; MAP@10
summarizes how consistently relevant documents appear throughout the top-10
ranking; MRR@10 focuses only on the first relevant hit. Reading the metrics
together is therefore important: an improvement in MRR@10 without improvement
in NDCG@10 or MAP@10 means that the system often finds one relevant document
earlier, but does not necessarily produce a better ranked list overall.

The main pattern is that legal nuggets help most when the query and the
extracted representation describe the same kind of legal unit. In JUA-Juris,
queries are ementas, i.e., summaries of legal theses. The nugget prompt was
also designed to extract legal theses from jurisprudence. This alignment
turns retrieval into a comparison between thesis-like queries and thesis-like
document representations, reducing the noise introduced by long
jurisprudential documents.

The JurisTCU result reinforces this interpretation. Even though the nugget
index was partly transferred from JUA-Juris and completed for missing
documents, the representation remains jurisprudence-oriented. It improves
NDCG@10, MAP@10, and MRR@10 over full-document embeddings,
indicating that the extracted theses capture useful retrieval evidence.

By contrast, NormasTCU and BR-TaxQA are not purely jurisprudential thesis
retrieval tasks. NormasTCU is especially informative because nugget retrieval
improves MRR@10 but lowers NDCG@10 and MAP@10. This indicates an early-hit
effect: nuggets can help surface at least one relevant normative document
sooner, but the remaining top-ranked results are less consistently useful than
those produced by full-document retrieval. One possible explanation is that
normative relevance often depends on broader regulatory context, definitions,
exceptions, or the relationship among provisions. Extracting isolated nuggets
may preserve some directly matchable statements while weakening the contextual
signals needed to rank several relevant norms well. BR-TaxQA shows an even
clearer case where nugget extraction is misaligned with the task: tax questions
and answers often depend on complete explanatory context, so full-document
answer representations remain stronger across NDCG@10, MAP@10, and MRR@10.

These findings point to a more precise claim: nugget retrieval should not be
treated as a generic replacement for full-document retrieval. Instead, it is a
task-specific representation strategy. It is particularly promising when the
information need is granular, legal-argumentative, and aligned with the
prompted extraction schema.

\subsection{Embedding Model Ablation}

We also run an ablation with stronger open-weight embedding models to examine
whether nugget retrieval remains useful when the embedding model changes.
Table~\ref{tab:embedding_ablation} compares full-document retrieval from the
JUA leaderboard with nugget retrieval using the same embedding model. We
consider two general Qwen3 embedding models \cite{zhang2025qwen3embedding}
and two JUA-adapted 4B models
\cite{pereira2026domainadaptivedenseretrievalbrazilian}.

\begin{table}[ht]
\centering
\caption{Embedding model ablation on jurisprudential datasets at cutoff 10.}
\label{tab:embedding_ablation}
\small
\setlength{\tabcolsep}{2.5pt}
\begin{tabular}{@{}lllrrr@{}}
\toprule
Dataset & Model & Approach & NDCG & MAP & MRR \\
\midrule
JUA-Juris & jua-4B-mixed & Full & \textbf{0.29044} & \textbf{0.23056} & \textbf{0.23026} \\
JUA-Juris & jua-4B-mixed & Nuggets & 0.24561 & 0.19034 & 0.19034 \\
JUA-Juris & jua-4B-legal-only & Full & \textbf{0.29371} & \textbf{0.23336} & \textbf{0.23313} \\
JUA-Juris & jua-4B-legal-only & Nuggets & 0.25398 & 0.19645 & 0.19645 \\
JUA-Juris & Qwen3-4B & Full & 0.19892 & 0.15181 & 0.15188 \\
JUA-Juris & Qwen3-4B & Nuggets & \textbf{0.22606} & \textbf{0.17354} & \textbf{0.17354} \\
JUA-Juris & Qwen3-8B & Full & 0.21719 & 0.16993 & 0.17002 \\
JUA-Juris & Qwen3-8B & Nuggets & \textbf{0.23499} & \textbf{0.18070} & \textbf{0.18070} \\
\midrule
JurisTCU & jua-4B-mixed & Full & \textbf{0.36337} & 0.16956 & \textbf{0.64069} \\
JurisTCU & jua-4B-mixed & Nuggets & 0.33752 & \textbf{0.17977} & 0.58679 \\
JurisTCU & jua-4B-legal-only & Full & \textbf{0.37520} & \textbf{0.17935} & \textbf{0.64986} \\
JurisTCU & jua-4B-legal-only & Nuggets & 0.32963 & 0.17422 & 0.57892 \\
JurisTCU & Qwen3-4B & Full & 0.31094 & 0.13752 & \textbf{0.58815} \\
JurisTCU & Qwen3-4B & Nuggets & \textbf{0.32423} & \textbf{0.16550} & 0.55645 \\
JurisTCU & Qwen3-8B & Full & \textbf{0.33175} & 0.14897 & 0.61106 \\
JurisTCU & Qwen3-8B & Nuggets & 0.32852 & \textbf{0.17115} & \textbf{0.62319} \\
\bottomrule
\end{tabular}
\end{table}

This ablation qualifies the main result. Nugget retrieval improves the
general Qwen3 embedding models on JUA-Juris, suggesting that granular legal
thesis representations can compensate for the lack of task-specific
adaptation. However, it does not outperform the JUA-adapted 4B models in the
full-document setting. On JurisTCU, nuggets often improve MAP@10 for the
general models, but this improvement does not consistently translate into
higher NDCG@10 or MRR@10.
Thus, the benefit of nugget retrieval depends not only on the document genre
and query type, but also on whether the embedding model has already learned a
strong document-level representation for the benchmark.

\subsection{Limitations}

This study has some limitations. First, nugget generation can be costly,
depending on the size of the search collection and on the quality and price of
the generation model. Future work should evaluate smaller or open-source
models for nugget extraction, especially in settings where the corpus is large
or frequently updated. Second, the document-level ranking is obtained with
simple aggregation strategies over retrieved nuggets. More sophisticated
scoring methods could combine nugget evidence, full-document scores, and
reranking models. Third, the evaluation is based on retrieval metrics from
existing benchmarks. These metrics are appropriate for comparing ranking
systems, but they do not directly measure how useful the retrieved legal
evidence is for a practitioner. Future work should therefore include
qualitative analysis of retrieved nuggets and their role in legal reasoning.
Finally, the experiments cover four Brazilian legal datasets and a limited set
of embedding models, so the conclusions should be tested on additional legal
collections and retrieval settings.

\section{Conclusion}

This paper investigated whether legal nugget extraction can improve dense
retrieval over long Brazilian jurisprudential texts. The proposed pipeline
extracts atomic legal theses from source documents, indexes these nuggets with
embeddings, retrieves nugget-level evidence for each query, and aggregates the
retrieved evidence back to document rankings.

The results support a qualified answer. Nugget retrieval improves retrieval on
the two jurisprudential datasets evaluated with \texttt{text-embedding-3-small},
especially when queries are themselves close to legal theses. However, the
approach is not uniformly better than full-document retrieval: it underperforms
on NormasTCU and BR-TaxQA, and the embedding-model ablation shows that strong
JUA-adapted retrievers remain better in the full-document setting. At the same
time, nuggets improve general Qwen3 embedding models on JUA-Juris and improve
MAP@10 in some JurisTCU comparisons, suggesting that granular representations
can be useful when the embedding model has not already learned a strong
document-level representation for the target benchmark.

Taken together, these findings suggest that legal nuggets should be viewed as
a task-sensitive representation strategy rather than a universal replacement
for full-document search. Their value depends on the alignment between the
prompt, the legal genre, the query formulation, and the embedding model used
for retrieval. Future work should study hybrid retrieval strategies that
combine nugget and full-document evidence, better aggregation methods,
cost-efficient nugget generation with smaller or open-source models, and
qualitative evaluation of the faithfulness and legal usefulness of retrieved
nuggets.

\section*{Ethics Statement}
\label{sec:ethical_considerations}

This work evaluates retrieval methods over existing legal benchmark datasets
and does not introduce new personal data collection or user-facing legal
decision systems. Nevertheless, legal retrieval is a high-stakes domain:
retrieved documents may influence legal research, institutional work, or
professional judgment. Nugget-based retrieval can make relevant legal theses
easier to find, but it can also omit context, surface incomplete propositions,
or overemphasize fragments that should be interpreted together with the full
decision. For this reason, the method should be viewed as a research tool for
retrieval and evidence discovery, not as a substitute for legal analysis by
qualified professionals.

\section*{Declaration of Generative AI Use}
\label{sec:generative_ai_usage}

The authors used generative AI tools to support language editing and grammar
correction during the preparation of this manuscript. Generative AI tools were
not used to generate experimental results, fabricate data, or perform the
quantitative evaluation reported in this paper. All scientific claims, system
descriptions, experimental results, and final text were reviewed, verified, and
approved by the authors, who remain fully responsible for the content of the
submission.
 
\bibliographystyle{sbc}
\bibliography{sbc-template}

\end{document}